\documentclass[letterpaper]{article} 
\usepackage[preprint]{aaai2027}
\usepackage[hyphens]{url}  
\usepackage{graphicx} 
\usepackage{natbib}  
\usepackage{caption} 
\usepackage{algorithm}
\usepackage{algorithmic}
\usepackage{amssymb}
\usepackage{multirow}
\usepackage{newfloat}
\usepackage{listings}
\DeclareCaptionStyle{ruled}{labelfont=normalfont,labelsep=colon,strut=off} 
\floatstyle{ruled}
\newfloat{listing}{tb}{lst}{}
\floatname{listing}{Listing}

\usepackage{booktabs}
\usepackage{amsmath}

\title{DeALR: Decoupled Audio-Latent Optimization with Acoustic Backtracking for Test-Time Reasoning}
\author{
    Haoyu He\textsuperscript{\rm 1}\equalcontrib,
    Yan Rong\textsuperscript{\rm 1}\equalcontrib,
    Li Liu\textsuperscript{\rm 1}\corresponding
}
\affiliations{

    \textsuperscript{\rm 1}HKUST(GZ)\\
    hhe152@connect.hkust-gz.edu.cn, yrong854@connect.hkust-gz.edu.cn, avrillliu@hkust-gz.edu.cn
}

\begin{document}

\maketitle

\begin{abstract}
Large audio-language models (LALMs) typically encode audio once and conduct subsequent multi-step reasoning in a language-dominated space, weakening access to fine-grained acoustic evidence. Although recent multimodal approaches have explored re-engaging perceptual inputs during reasoning, whether a frozen LALM can actively revisit crucial acoustic evidence in the latent space at test time remains underexplored. 
To bridge this gap, we introduce Decoupled Audio Latent Reasoning (DeALR), a novel test-time latent reasoning framework that shifts the paradigm from text-dominated deliberation to active audio-latent interaction. Inspired by the human cognitive process of "backtracking" to specific acoustic cues during complex listening tasks, DeALR establishes a dynamic interaction between learnable thought tokens and acoustic segments. 
Specifically, we introduce Span-Level Acoustic Backtracking, which traces initial attention distributions to dynamically isolate contiguous, key acoustic spans. 
Furthermore, we propose Decoupled Audio-Latent Optimization (DALO), a test-time reinforcement learning algorithm that jointly refines thought vectors and acoustic patches via confidence-guided policy gradients. Crucially, DALO explicitly decouples trajectory search from response generation by reinstating uncorrupted, high-fidelity audio embeddings prior to text synthesis, entirely eliminating optimization-induced distribution drift while retaining the navigation of the refined thoughts. 
Comprehensive experiments on benchmarks demonstrate that DeALR consistently outperforms prior test-time optimization approaches, delivering substantial improvements in both confidence calibration and reasoning accuracy.
\end{abstract}
\section{Introduction}

Audio-Language Models have emerged as a powerful paradigm for bridging human auditory perception with linguistic intelligence. Unlike traditional speech-to-text systems, modern audio reasoning demands that models not only transcribe words, but also comprehend complex acoustic scenes, identify paralinguistic cues, and perform higher-level semantic deduction over extended audio streams. As benchmarks like MMAR~\cite{ma2026mmar} and MMAU~\cite{sakshi2025mmau} push the boundaries of the field, the requirement for deep, multi-step deliberation over multimodal inputs has become a core challenge in audio intelligence.

To tackle such complex reasoning tasks, Large Language Models traditionally rely on Chain-of-Thought (CoT) prompting~\cite{wei2022chain}, which breaks down intricate problems into explicit intermediate linguistic steps. However, generating explicit textual tokens at every reasoning step introduces immense computational overhead and inference latency. To alleviate this bottleneck, recent advances have shifted towards latent reasoning, which compresses intermediate thoughts into continuous hidden vectors. Frameworks like Coconut~\cite{hao2024training} and SoftCoT~\cite{xu2025softcot} have demonstrated that models can perform complex multi-step deliberation entirely within the latent space, maintaining high accuracy while significantly accelerating inference.

Despite its success in the text and visual domain~\cite{ye2025thinking, liu2025reasoning}, transferring latent reasoning to the audio domain reveals fundamental architectural flaws. In standard ALM architectures, the rich, continuous acoustic signal is typically encoded only once during the initial forward pass, while subsequent latent deliberation unfolds almost exclusively within the linguistic subspace. This gives rise to two critical challenges: (1) Acoustic evidence fading: fine-grained acoustic cues are progressively overwhelmed by dominant language priors, forcing the model to rely on suboptimal textual compensation rather than the actual audio signals. (2) Lack of adaptive audio-latent interaction: existing frozen models lack a mechanism to actively backtrack and dynamically re-engage with acoustic evidence in the latent space, leaving the crucial question of dynamic audio-latent reasoning at test time entirely unexplored. 

\begin{figure}[t]
    \centering
    \includegraphics[width=\columnwidth]{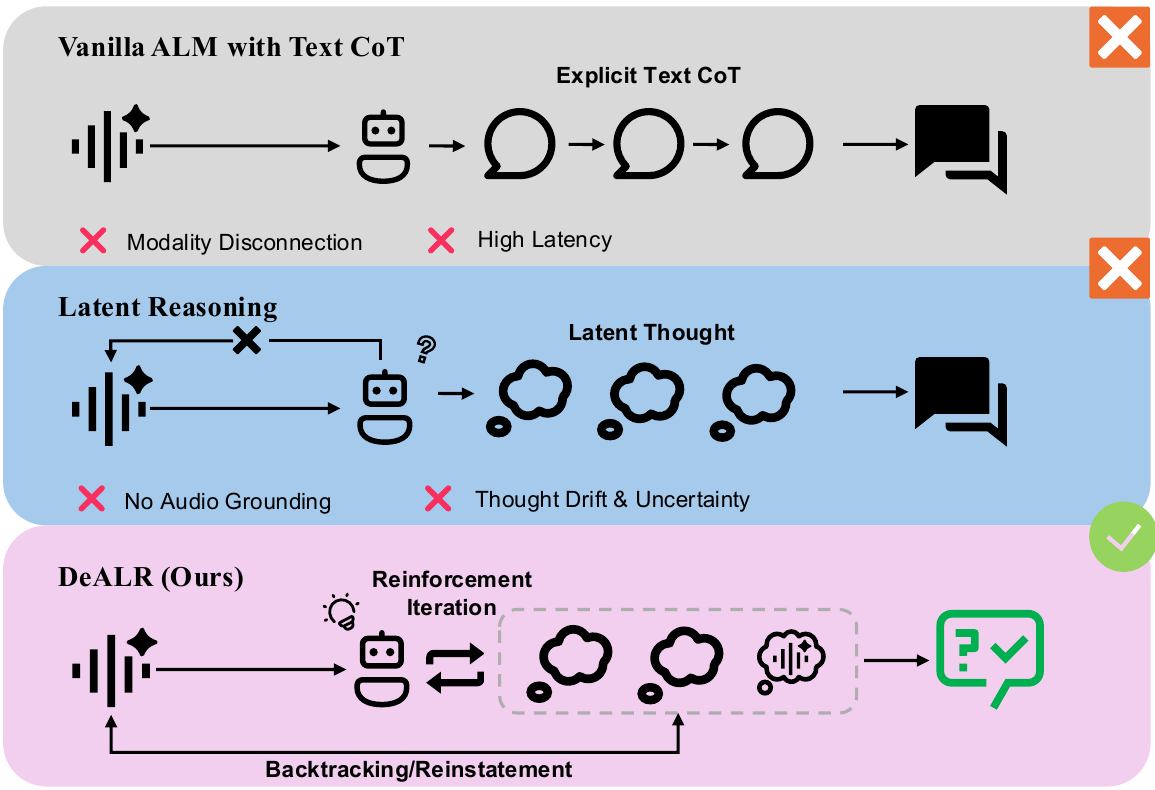}
    \caption{Comparison of reasoning paradigms.
(a) Vanilla ALM with text-based CoT.
(b) Latent reasoning without acoustic revisiting.
(c) DeALR dynamically backtracks to salient acoustic spans and performs audio-latent co-optimization.}
    \label{fig:paradigm}
\end{figure}

To address these challenges, we introduce DeALR (\textbf{De}coupled \textbf{A}udio \textbf{L}atent \textbf{R}easoning), a novel test-time latent reasoning framework that shifts the paradigm from text-dominated deliberation to active audio-latent interaction. As shown in Figure~\ref{fig:paradigm}, unlike prior paradigms that either rely on text-based CoT or allow acoustic evidence to fade, DeALR actively revisits audio spans during latent deliberation. Our approach is inspired by the human cognitive process during complex listening tasks: when faced with ambiguous or intricate audio, humans instinctively "backtrack" to specific acoustic segments to verify and refine their understanding. Mirroring this mechanism, DeALR treats both intermediate thought tokens and acoustic representations as dynamic parameters, jointly optimizing them at test time without altering the frozen weights of the base model. Concretely, DeALR comprises two synergistic components, each designed to tackle one of the aforementioned challenges. For the first challenge, we propose Span-Level Acoustic Backtracking, which traces the model's initial attention distributions to dynamically isolate contiguous, key acoustic spans and interleaves them with latent thought vectors, thereby reinjecting fine-grained acoustic evidence directly into the reasoning stream. For the second challenge, we propose Decoupled Audio-Latent Optimization (DALO), a test-time reinforcement learning algorithm that jointly refines thought vectors and acoustic patches via confidence-guided policy gradients. Crucially, DALO explicitly decouples trajectory search from response generation, restoring uncorrupted acoustic embeddings prior to text synthesis and eliminating the optimization-induced distribution drift.

In summary, our main contributions are as follows:
\begin{itemize}
    \item We formulate test-time audio latent reasoning as active acoustic evidence revisiting. To the best of our knowledge, \textsc{DeALR} is the first training-free framework to combine latent search with span-level acoustic revisiting in a frozen LALM, shifting audio from static conditioning to an active source of evidence.
    \item We introduce Span-Level Acoustic Backtracking, which converts thought-to-audio attention into a compact set of contiguous temporal spans and routes those spans back into the latent reasoning sequence.
    \item We develop \textsc{DALO}, a confidence-guided joint search over latent thoughts and selected acoustic representations. Its decoupled decoding strategy reinstates original audio embeddings at the selected locations, mitigating search-induced representation drift.
    \item Comprehensive evaluations on MMAR and MMAU validate the superiority of DeALR, demonstrating its ability to resolve critical failure points caused by dominant language priors in existing audio-language models.
\end{itemize}
\section{Related Work}

\subsection{Multimodal Reasoning}
Advanced reasoning in Large Language Models (LLMs) has evolved from explicit textual rationales via Chain-of-Thought (CoT) prompting~\cite{wei2022chain, kojima2022large} to sophisticated multimodal paradigms that bridge perception and inference. In multimodal settings, early approaches primarily perform reasoning entirely within the text space after a single round of visual or sensory encoding~\cite{lu2022learn, zhang2023multimodal, tang2024salmonn}. While straightforward, such text-mediated architectures often suffer from severe grounding errors and dominant language biases, as the rigid separation between perception and reasoning leads to hallucinations and acoustic/visual evidence fading. To mitigate these issues, recent vision-language models adopt a ``thinking-with-images'' paradigm~\cite{zheng2026deepeyes, gao2025interleaved}, dynamically manipulating visual inputs during inference via region zooming, visual editing, or intermediate cue generation. However, these approaches introduce massive computational overhead, remain highly vulnerable to unstable tool invocations, and fail to accommodate the continuous, sequential nature of auditory signals~\cite{deshmukh2023pengi}.

\subsection{Latent Space Reasoning}
Latent space reasoning represents intermediate thoughts as continuous hidden vectors instead of discrete CoT tokens, reducing inference latency while preserving reasoning quality~\cite{shen2026llmdoctor, shi2025swireasoning, goyal2023think, zelikman2024quiet}. Most existing methods rely on offline training, such as Coconut~\cite{hao2024training} (using previous latent states as input embeddings) and SoftCoT~\cite{xu2025softcot} (projecting soft tokens via a lightweight module). ICoT~\cite{gao2025interleaved} avoids training entirely through parameter‑free selection. Despite their efficiency, these paradigms remain predominantly text‑centric and treat multimodal inputs as static, frozen representations.

\subsection{Reinforcement Learning in Test-Time Optimization}
Reinforcement Learning (RL) has long been utilized to enhance LLM reasoning capabilities. While traditional methods like RLHF~\cite{ouyang2022training} or GRPO~\cite{guo2025deepseek,shao2024deepseekmath} fine-tune model weights through heavy offline training phases, they demand substantial computational resources and lack per-instance adaptability. To achieve flexible, instance-level adaptation, recent studies have pivoted towards parameter-free test-time optimization (TTO)~\cite{li2025seek, yao2023tree, snell2024scaling} via policy gradients. For example, LTPO~\cite{ye2025thinking} treats latent thought vectors as dynamic parameters optimized at test time via a confidence-based reward signal. Extending this to the multimodal domain, DMLR~\cite{liu2025reasoning} introduces a dynamic visual injection strategy to retrieve task-relevant image patches alongside confidence-guided optimization. 
\section{Preliminary and Motivation}

\subsection{Preliminary}
\paragraph{Audio Reasoning.} Given a text query $\mathcal{Q} = (q_1, \ldots, q_L)$ and a continuous audio input $\mathcal{A}$, a standard Large Audio-Language Model (LALM) processes the audio through an encoder to yield sequence embeddings $\mathcal{Z} = (z_1, \ldots, z_T) \in \mathbb{R}^{T \times d}$, where $d$ is the hidden dimension and $T$ is the temporal length. Conventional naive LALMs directly predict the target text sequence $\mathcal{Y}$ conditioned solely on the audio and query context via $\pi_\theta(\mathcal{Y} \mid \mathcal{Z}, \mathcal{Q})$.

\paragraph{Audio Latent Reasoning.} However, complex audio understanding often requires test-time deliberation. To address this, an intermediate sequence of latent thinking tokens $\mathcal{T} = (t_1, \ldots, t_K)$ is introduced, reformulating the final response generation to be conditioned on the augmented context via $\pi_\theta(\mathcal{Y} \mid \mathcal{Z}, \mathcal{Q}, \mathcal{T})$. By incorporating $\mathcal{T}$, the model is enabled to perform structured reasoning within the continuous latent space before committing to the final textual answer.

\subsection{Motivation}

\paragraph{Do Conventional Methods Sufficiently Attend to Audio?}
To quantitatively evaluate whether model decisions are actively grounded in acoustic signals rather than dominated by textual priors, we introduce a diagnostic metric termed Audio Attention Ratio (AAR). Given an $L$-layer model with $H$ attention heads, we first obtain the head-averaged attention matrix $\mathbf{A}^{(l)} \in \mathbb{R}^{S \times S}$ at layer $l$, where $\mathbf{A}^{(l)}_{q, k} = \frac{1}{H} \sum_{h=1}^{H} \mathbf{A}^{(l, h)}_{q, k}$ represents the attention weight assigned from query position $q$ to key position $k$. The AAR for a query at position $q$ aggregated across all layers is defined as:
\begin{equation}
\text{AAR}(q) = \frac{\sum_{l=1}^{L} \sum_{k \in \mathcal{K}_{\text{audio}}} \mathbf{A}^{(l)}_{q, k}}{\sum_{l=1}^{L} \sum_{k \in \mathcal{K}_{\text{all}}} \mathbf{A}^{(l)}_{q, k}},
\end{equation}
where $\mathcal{K}_{\text{audio}}$ and $\mathcal{K}_{\text{all}}$ denote the set of audio key indices and total key indices, respectively.

\begin{figure}[htbp]
    \centering
    \includegraphics[width=\linewidth]{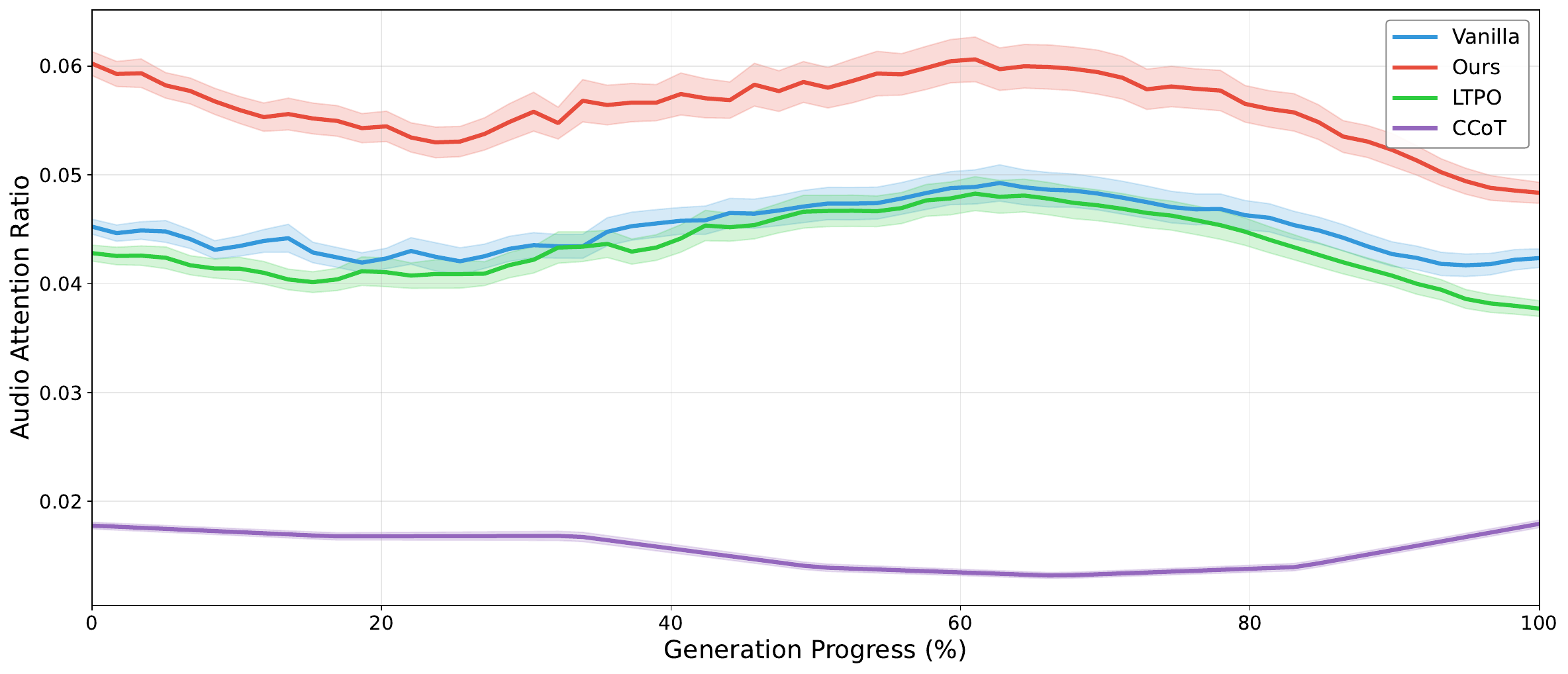}
    \caption{AAR comparison among Vanilla LALM, LTPO, CCoT, and Ours.}
    \label{fig:aar_comparison}
\end{figure}

As illustrated in Figure~\ref{fig:aar_comparison}, conventional vanilla LALMs and naive latent baselines maintain a low AAR across almost all query positions. This confirms that their latent dynamics and text decoders predominantly attend to textual context $\mathcal{Q}$ and prior language tokens while largely ignoring audio sequence tokens $\mathcal{Z}$. In contrast, our proposed framework significantly restores the AAR during reasoning, ensuring that the continuous reasoning sequence is actively grounded on acoustic features.

\begin{figure}[htbp]
    \centering
    \includegraphics[width=\linewidth]{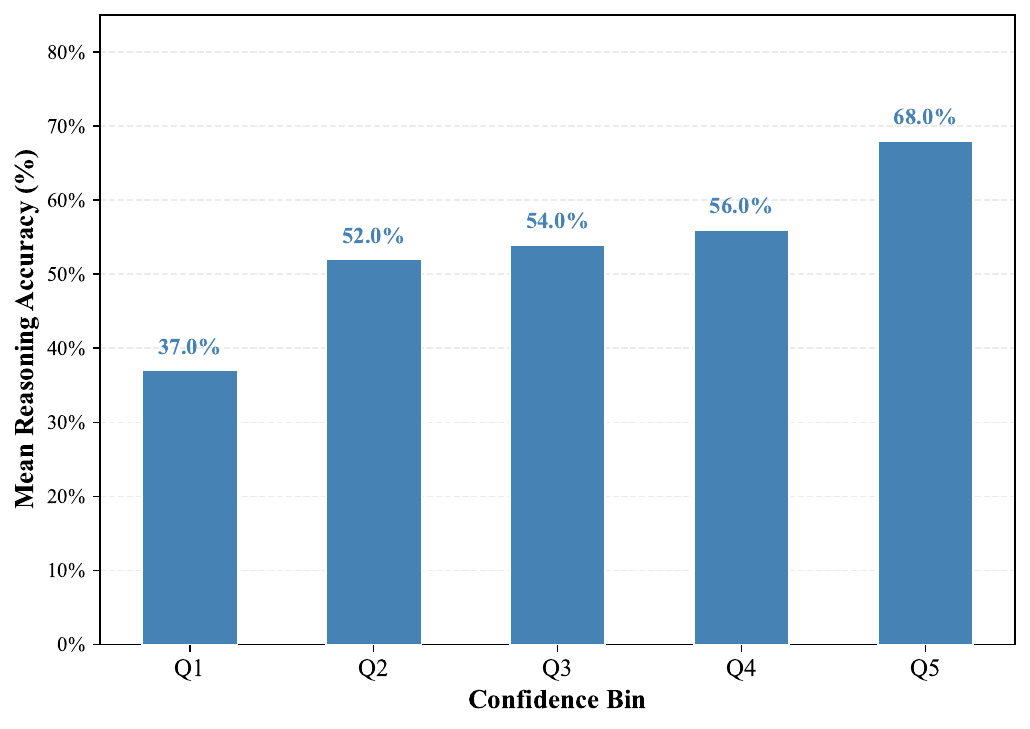}
    \caption{Correlation between reasoning accuracy and model confidence.}
    \label{fig:confidence_search}
\end{figure}

\paragraph{Can Confidence Guide Latent Search?}
Given the limitations of standard autoregressive generation, we hypothesize that test-time search within the continuous latent space can guide the model toward better reasoning paths. However, navigating the latent space requires an effective quality signal. We examine whether model confidence can serve as a reliable heuristic during test-time latent search. Therefore, we collected reasoning trajectories of Qwen2.5-Omni~\cite{xu2025qwen25omnitechnicalreport}. As illustrated in Figure~\ref{fig:confidence_search}, trajectories with higher confidence scores consistently correlate with correct reasoning pathways and stronger alignment with audio cues. This suggests that predictive confidence can guide latent tokens T toward promising paths without external supervision.

\section{Methodology}
\begin{figure*}[t] 
    \centering
    \includegraphics[width=\linewidth]{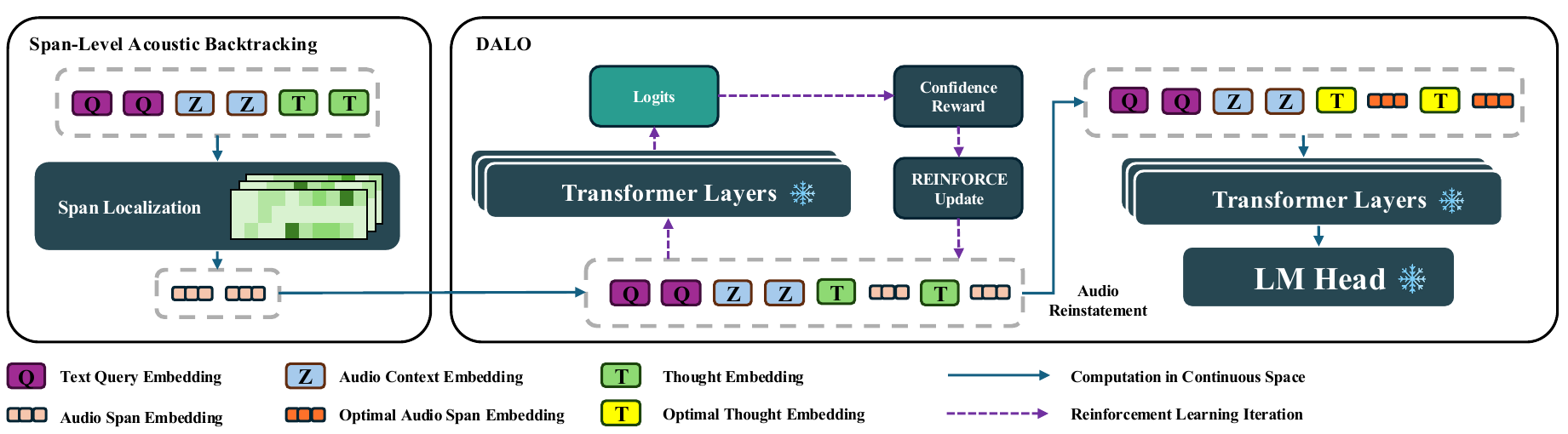}
    \caption{Overall architecture of DeALR. The framework consists of two main components: (1) Span-Level Acoustic Backtracking, which selects salient acoustic spans and interleaves them with thought tokens; (2) Decoupled Audio-Latent Optimization (DALO), which jointly searches optimal thought tokens and audio patches via confidence-guided policy gradient, then reinstates raw embeddings for final decoding.}
    \label{fig:framework}
\end{figure*}

The overall architecture of DeALR is illustrated in Figure~\ref{fig:framework}.

\subsection{Span-Level Acoustic Backtracking}

\subsubsection{Latent-Guided Acoustic Span Localization}
To enable the model to actively locate essential acoustic evidence, we extract salient acoustic spans guided by the initial latent thought tokens. We augment the input prompt $\mathcal{Q}$ with $K$ initial latent thought tokens $\mathcal{T}^{(0)} = (\tau_1, \ldots, \tau_K)$, which are initialized using dedicated special tokens, and execute a forward pass through $\pi_\theta$. We extract the attention matrix $\mathbf{A} \in \mathbb{R}^{K \times T}$, where $\mathbf{A}_{i,j}$ represents the attention weight from the $i$-th thought token $\tau_i$ to the $j$-th audio token $z_j$. The aggregated acoustic saliency score $\mathbf{s}_j$ for each audio position $j$ is obtained by averaging across all thought tokens:
\begin{equation}
\mathbf{s}_j = \frac{1}{K} \sum_{i=1}^{K} \mathbf{A}_{i,j}, \quad \forall j \in \{1, \ldots, T\}.
\end{equation}

To mitigate pointwise attention noise, we apply 1D sliding window average pooling with kernel size $w_s$ to smooth the saliency sequence:
\begin{equation}
\tilde{\mathbf{s}}_j = \frac{1}{w_s} \sum_{t = j - \lfloor w_s/2 \rfloor}^{j + \lfloor w_s/2 \rfloor} \mathbf{s}_t.
\end{equation}

Next, we perform 1D Temporal Non-Maximum Suppression (NMS) to isolate $K$ temporally distinct anchor positions. Candidate positions are sorted by $\tilde{\mathbf{s}}_j$ in descending order. We iteratively select the candidate with the highest score as an anchor $p_k$ and suppress surrounding candidates within a radius $r$, yielding $K$ anchor points $\{p_1, \ldots, p_K\}$. Finally, each anchor $p_k$ is expanded into a contiguous acoustic span $\mathcal{C}_k$ of width $w$:
\begin{equation}
\mathcal{C}_k = \left( z_{p_k - \lfloor w/2 \rfloor}, \ldots, z_{p_k}, \ldots, z_{p_k + \lfloor w/2 \rfloor} \right) \in \mathbb{R}^{w \times d}.
\end{equation}

\subsubsection{Span-Thought Interleaving}
To allow the model to alternate between implicit reasoning and explicit acoustic verification, we interleave the extracted continuous audio spans $\mathcal{C}_k$ with the latent thought tokens $\tau_k$. Specifically, each thought token $\tau_k$ is directly followed by its corresponding acoustic span $\mathcal{C}_k$, yielding the structured input sequence:
\begin{equation}
\mathcal{X} = \big[\mathcal{Q}_{\text{prefix}}, \; \tau_1, \mathcal{C}_1, \; \tau_2, \mathcal{C}_2, \; \ldots, \; \tau_K, \mathcal{C}_K, \; \mathcal{Q}_{\text{suffix}}\big].
\end{equation}
We denote $\mathcal{P} = \{\mathcal{C}_1, \ldots, \mathcal{C}_K\} \in \mathbb{R}^{(K \cdot w) \times d}$ as the concatenated tensor of selected audio patch embeddings.

\subsection{Decoupled Audio-Latent Optimization}
\label{sec:dalo}

To maximize test-time reasoning alignment, DALO jointly optimizes thought tokens $\mathcal{T}$ and audio patch embeddings $\mathcal{P}$ via policy gradient search, followed by a reinstatement step to preserve raw acoustic semantics.

\subsubsection{Joint Latent-Acoustic Search and Update}
During the $M$-step search phase, we explore candidate reasoning trajectories by perturbing both thought tokens $\mathcal{T}$ and audio patch embeddings $\mathcal{P}$ with isotropic Gaussian noise, whose variance decays exponentially over time ($\sigma_m = \sigma_0 \cdot \gamma^m, \gamma \in (0, 1)$).

Without ground-truth textual supervision at test time, we measure reasoning quality via the model's output confidence over thought tokens. For thought token $\tau_i$, its individual confidence $C(\tau_i)$ is computed using the average negative log-probability of its Top-$k$ predicted tokens~\cite{fu2026deep}:
\begin{equation}
C(\tau_i) = -\frac{1}{k} \sum_{v \in \text{Top}_k(P_i)} \log P_i(v),
\end{equation}
where $P_i = \pi_\theta(\cdot \mid \mathcal{X}, <\tau_i)$. The scalar intrinsic reward $\mathcal{R}^{(m)}$ at step $m$ is defined as the mean confidence across all $K$ thought tokens:
\begin{equation}
\mathcal{R}^{(m)} = \frac{1}{K} \sum_{i=1}^{K} C(\tau_i).
\end{equation}

Using the REINFORCE formulation under Gaussian exploration, the joint parameter updates are derived as:
\begin{equation}
\mathcal{T}^{(m+1)} = \mathcal{T}^{(m)} + \eta \cdot \mathcal{R}^{(m)} \cdot \frac{\epsilon_{\mathcal{T}}^{(m)}}{\sigma_m^2},
\end{equation}
\begin{equation}
\mathcal{P}^{(m+1)} = \mathcal{P}^{(m)} + \eta \cdot \mathcal{R}^{(m)} \cdot \frac{\epsilon_{\mathcal{P}}^{(m)}}{\sigma_m^2},
\end{equation}
where $\eta$ is the test-time learning rate. Across $M$ exploration steps, we track the state $m^*$ that maximizes $\mathcal{R}^{(m)}$, yielding the optimal thought tokens $\mathcal{T}^*$ and optimal anchor indices $\{p_1^*, \ldots, p_K^*\}$.


\subsubsection{Audio Embedding Reinstatement and Final Decoding}
While allowing updates to patch embeddings $\mathcal{P}$ during search facilitates smooth joint landscape navigation, directly feeding perturbed embeddings $\mathcal{P}^*$ into the final text generation stage can induce severe distribution drift and semantic degradation. 

To overcome this, DALO explicitly decouples the test-time search navigation phase from the final autoregressive decoding phase. Specifically, upon concluding the exploration steps, we retain the optimal latent thought tokens $\mathcal{T}^*$ alongside their dynamically tracked temporal anchor locations $\{p_1^*, \ldots, p_K^*\}$, while completely discarding any perturbed intermediate audio representations. To bridge the optimal reasoning path with authentic acoustic evidence, we reinstate the uncorrupted, high-fidelity audio embeddings $\mathcal{Z}_{\mathcal{C}_k^*}$, which are extracted directly from the original audio sequence $\mathcal{Z}$ at those optimal anchor locations. 

The realigned input sequence for final decoding is formulated as:
\begin{equation}
\mathcal{X}_{\text{final}} = \big[\mathcal{Q}_{\text{prefix}}, \; \tau_1^*, \mathcal{Z}_{\mathcal{C}_1^*}, \; \tau_2^*, \mathcal{Z}_{\mathcal{C}_2^*}, \; \ldots, \; \tau_K^*, \mathcal{Z}_{\mathcal{C}_K^*}, \; \mathcal{Q}_{\text{suffix}}\big].
\end{equation}
Finally, $\mathcal{X}_{\text{final}}$ is passed to the LLM backend to generate the response via standard autoregressive decoding:
\begin{equation}
\mathcal{Y}^* = \arg\max_{\mathcal{Y}} \prod_{n=1}^{N} \pi_\theta(y_n \mid \mathcal{Y}_{<n}, \mathcal{X}_{\text{final}}).
\end{equation}

\begin{algorithm}[tb]
\caption{Decoupled Audio-Latent Optimization}
\label{alg:dalo}
\textbf{Input}: text query $\mathcal{Q}$, initial thought tokens $\mathcal{T}^{(0)}$, initial audio patch embeddings $\mathcal{P}^{(0)}$, audio sequence embeddings $\mathcal{Z}$, frozen LALM $\pi_\theta$\\
\textbf{Output}: $\mathcal{Y}^*$
\begin{algorithmic}[1]
\STATE \# \textit{Phase 1: Initialization}
\STATE Initialize $\mathcal{R}^* \leftarrow -\infty$, $\mathcal{T}^* \leftarrow \mathcal{T}^{(0)}$, $\mathcal{C}^* \leftarrow \text{Initial Spans}$.
\STATE \# \textit{Phase 2: Joint Latent-Acoustic Search and Update}
\FOR{$m = 0$ \textbf{to} $M-1$}
    \STATE $\sigma_m \leftarrow \sigma_0 \cdot \gamma^m$
    \STATE $\epsilon_{\mathcal{T}} \sim \mathcal{N}(0, \sigma_m^2 \mathbf{I})$, $\epsilon_{\mathcal{P}} \sim \mathcal{N}(0, \sigma_m^2 \mathbf{I})$
    \STATE $\tilde{\mathcal{T}} \leftarrow \mathcal{T}^{(m)} + \epsilon_{\mathcal{T}}$, $\tilde{\mathcal{P}} \leftarrow \mathcal{P}^{(m)} + \epsilon_{\mathcal{P}}$
    \STATE Compute reward $\mathcal{R}^{(m)} = \text{Confidence}(\tilde{\mathcal{T}})$
    \IF{$\mathcal{R}^{(m)} > \mathcal{R}^*$}
        \STATE $\mathcal{R}^* \leftarrow \mathcal{R}^{(m)}$, $\mathcal{T}^* \leftarrow \tilde{\mathcal{T}}$
        \STATE $\mathcal{C}^* \leftarrow \text{1D-NMS-Spans}(\tilde{\mathcal{T}}, \mathcal{Z})$
    \ENDIF
    \STATE $\mathcal{T}^{(m+1)} \leftarrow \mathcal{T}^{(m)} + \eta \mathcal{R}^{(m)} \frac{\epsilon_{\mathcal{T}}}{\sigma_m^2}$
    \STATE $\mathcal{P}^{(m+1)} \leftarrow \mathcal{P}^{(m)} + \eta \mathcal{R}^{(m)} \frac{\epsilon_{\mathcal{P}}}{\sigma_m^2}$
\ENDFOR
\STATE \# \textit{Phase 3: Audio Embedding Reinstatement and Decoding}
\STATE $\mathcal{X}_{\text{final}} \leftarrow \big[\mathcal{Q}_{\text{prefix}}, \tau_1^*, \mathcal{Z}_{\mathcal{C}_1^*}, \dots, \tau_K^*, \mathcal{Z}_{\mathcal{C}_K^*}, \mathcal{Q}_{\text{suffix}}\big]$
\STATE Decode: $\mathcal{Y}^* \leftarrow \arg\max \pi_\theta(\cdot \mid \mathcal{X}_{\text{final}})$
\STATE \textbf{return} $\mathcal{Y}^*$
\end{algorithmic}
\end{algorithm}


\section{Experiments}

\subsection{Experimental Setup}

\paragraph{Datasets and Benchmarks.} We evaluate our method on three audio understanding and reasoning benchmarks: MMAR~\cite{ma2026mmar}, MMAU~\cite{sakshi2025mmau}, AudioMCQ~\cite{he2025measuring}.

\paragraph{Baselines.}
We compare against representative paradigms:
(1) Vanilla: Zero-shot direct generation;
(2) CCoT~\cite{mitra2024compositional} : Standard explicit text-based reasoning;
(3) ICoT~\cite{gao2025interleaved} : Interleaved textual reasoning with static multimodal features;
(4) LTPO~\cite{ye2025thinking} : Test-time latent optimization operating exclusively on text thought vectors;
(5) DMLR~\cite{liu2025reasoning} : Confidence-guided latent policy optimization with dynamic visual feature interleaving;

\paragraph{Implementation Details.}
We adopt Qwen2.5-Omni-7B~\cite{xu2025qwen25omnitechnicalreport} and Qwen2-Audio-Instruct~\cite{chu2024qwen2audiotechnicalreport} as our base model. All original LLM parameters remain frozen. During test-time optimization, we optimize only $32$ learnable vectors, comprising $8$ latent thought tokens ($\mathbf{\Theta} \in \mathbb{R}^{8 \times D}$) and $24$ patch vectors ($\mathbf{P} \in \mathbb{R}^{24 \times D}$, corresponding to $N=8$ chunks with window width $w=3$). 
Acoustic selection is guided by 1D temporal NMS with an inhibition radius $r=30$ frames on the all-layer mean attention matrix. REINFORCE optimization runs for $N_{\text{steps}}=15$ iterations per sample using learning rate $\eta=0.005$, initial exploration noise $\sigma_0=20$ with a decay factor of $0.95$. All experiments are conducted on 8 NVIDIA GeForce RTX 3090 (24 GB) GPUs.

\subsection{Comparison with Test-Time Reasoning Methods}

A comparison of question-answering accuracy across methods and benchmarks is presented in Table~\ref{tab:main_results}.

\begin{table*}[t]
\centering
\small
\caption{Comparison of different reasoning methods and \textsc{DeALR} across various audio benchmarks under multiple backbone models. All metrics are reported in Accuracy (\%). Results demonstrate consistent performance gains over both base models and existing test-time optimization approaches. Best and second-best results are in bold and underlined, respectively. (LP-MTT: LP-MusicCaps-MTT, Speech: SpeechCraft)}
\label{tab:main_results}
\resizebox{\linewidth}{!}{
\setlength{\tabcolsep}{3.5pt} 
\begin{tabular}{ll|c|c|cccccccc}
\toprule
\multirow{2}{*}{\textbf{Model}} & \multirow{2}{*}{\textbf{Method}} & \multirow{2}{*}{\textbf{MMAR}} & \multirow{2}{*}{\textbf{MMAU}} & \multicolumn{8}{c}{\textbf{AudioMCQ}} \\
\cmidrule(lr){5-12}
& & & & AudioCaps & Clotho & CompA-R & LP-MTT & MusicCaps & Speech & Tacos & Avg \\
\midrule
\multirow{6}{*}{Qwen2.5-Omni-7B} 
 & Vanilla & 57.7 & 70.2 & 83.4 & \underline{81.4} & 75.8 & \underline{94.3} & 81.0 & 79.0 & 42.3 & 78.1 \\
 & CCoT & 57.8 & \underline{72.7} & 84.6 & 80.4 & \textbf{78.0} & 82.4 & \underline{86.6} & 80.4 & \textbf{46.5} & \underline{79.7} \\
 & ICoT & 56.0 & 38.0 & 25.3 & 20.6 & 25.0 & 30.0 & 23.2 & 31.1 & 15.0 & 27.8 \\
 & LTPO & \underline{58.6} & 71.5 & 80.4 & 69.6 & 75.7 & 90.7 & 73.9 & 75.6 & 40.0 & 75.3 \\
 & DMLR & 58.2 & 67.3 & \underline{85.9} & 78.7 & 74.6 & \textbf{95.7} & \textbf{82.5} & \underline{81.4} & 32.6 & 77.4 \\
 & \textbf{\textsc{DeALR} (Ours)} & \textbf{60.3} {\scriptsize $\uparrow 2.6\%$} & \textbf{73.1} {\scriptsize $\uparrow 2.9\%$} & \textbf{88.4} & \textbf{83.5} & \underline{76.2} & 91.1 & 78.2 & \textbf{81.5} & \underline{44.1} & \textbf{80.5} {\scriptsize $\uparrow 2.4\%$} \\
\midrule
\multirow{6}{*}{Qwen2-Audio-Instruct} 
     & Vanilla & 36.8 & 58.4 & \underline{87.0} & \underline{78.9} & \underline{73.3} & \underline{92.2} & 82.4 & \underline{79.0} & \underline{28.2} & \underline{77.7} \\
 & CCoT & \textbf{41.1} & \underline{59.1} & 81.0 & 84.6 & 66.9 & 76.5 & \underline{85.7} & 68.7 & 23.9 & 69.0 \\
 & ICoT & 27.3 & 45.0 & 29.9 & 29.9 & 20.2 & 24.8 & 31.0 & 23.5 & 5.8 & 24.5 \\
 & LTPO & 37.9 & 58.4 & 66.9 & 57.2 & 59.9 & 61.0 & 59.2 & 56.8 & 27.2 & 58.2 \\
 & DMLR & -- & -- & -- & -- & -- & -- & -- & -- & -- & -- \\
 & \textbf{\textsc{DeALR} (Ours)} & \underline{38.1} {\scriptsize $\uparrow 1.3\%$} & \textbf{60.7} {\scriptsize $\uparrow 2.3\%$} & \textbf{89.0} & \textbf{79.9} & \textbf{80.1} & \textbf{95.3} & \textbf{85.9} & \textbf{80.4} & \textbf{36.9} & \textbf{80.6} {\scriptsize$\uparrow  2.9\%$} \\
\bottomrule
\end{tabular}
}
\end{table*}
\textsc{DeALR} consistently achieves superior accuracy across all datasets. Crucially, the performance of ICoT is severely degraded due to directly inserting audio embeddings into the autoregressive stream, which may disrupt the sequential generation. We also note that DMLR is excluded from our Qwen2-Audio evaluation due to the absence of suitable special tokens required to initialize its latent think tokens.


\subsection{Ablation Study}
To systematically quantify the individual contributions of each core module in \textsc{DeALR}, we conduct extensive ablation experiments across the MMAR and MMAU benchmarks. We categorize our ablations into three aspects: (1) Component-level breakdown of the \textsc{DeALR} pipeline, (2) Structural design of Span-Level Acoustic Backtracking, and (3) Search and update dynamics within DALO.

\paragraph{Analysis of Components.}
We isolate the core components of \textsc{DeALR} to investigate their individual contributions, with results summarized in Table~\ref{tab:ablation_component}. 
First, removing Audio Embedding Reinstatement yields the most catastrophic performance degradation. This empirical gap validates our core hypothesis: while continuous latent perturbation facilitates reasoning trajectory exploration, feeding noisy, drifted embeddings directly into the LLM decoder severely distorts acoustic semantics. Restoring uncorrupted acoustic features is therefore vital for high-fidelity generation.
Second, disabling Test-Time Latent Search causes a notable penalty, confirming that static zero-shot forward passes are insufficient for complex audio reasoning, and that confidence-guided deliberation is critical for navigating toward optimal thought paths.
Finally, eliminating Acoustic Span Localization, Span-Thought Interleaving, or Joint Optimization consistently erodes accuracy. These structural modules operate synergistically to ensure fine-grained cross-modal alignment and active perception-reasoning interaction.

\begin{table}[h]
\centering
\caption{Component-level ablation study of \textsc{DeALR} on MMAU.}
\label{tab:ablation_component}
\resizebox{\linewidth}{!}{
\begin{tabular}{lccc}
\toprule
\textbf{Framework Variant}  & \textbf{Acc(\%)} & \textbf{$\Delta$} \\
\midrule
\textbf{Full \textsc{DeALR} Framework}  & \textbf{73.1} & -- \\
\quad \textit{w/o} Audio Embedding Reinstatement & 68.4 & -4.7 \\
\quad \textit{w/o} Joint Optimization & 71.9 & -1.2 \\
\quad \textit{w/o} Span-Thought Interleaving & 71.7 & -1.4 \\
\quad \textit{w/o} Acoustic Span Localization & 71.5 & -1.6 \\
\quad \textit{w/o} Test-Time Latent Search ($M=0$) & 70.2 & -2.9 \\
\bottomrule
\end{tabular}
}
\end{table}

\subsubsection{Design Analysis of Span-Level Acoustic Backtracking.}
We examine the topological choices in acoustic span extraction by evaluating different localization strategies and span widths, as shown in Table~\ref{tab:ablation_backtracking}. Relying on the full audio context ($\mathcal{Z}$) or uniformly sampling acoustic spans yields suboptimal performance, primarily due to unconstrained temporal noise and the dilution of salient acoustic events. While pointwise Top-$K$ selection ($w=1$) improves precision by filtering out background noise, isolated frames fail to preserve local phonetic and acoustic continuity. In contrast, our 1D Temporal NMS strategy with expanded spans ($w=3$) achieves the best performance across both benchmarks.

\begin{table}[h]
\centering
\caption{Ablation on acoustic span localization strategies.}
\label{tab:ablation_backtracking}
\resizebox{\linewidth}{!}{
\begin{tabular}{llcc}
\toprule
\textbf{Strategy}  & \textbf{MMAU} \\
\midrule
Full Audio Context ($\mathcal{Z}$) & 70.7 \\
Uniform Acoustic Spans & 68.3 \\
Random Acoustic Spans & 71.6 \\
Pointwise Top-$K$ Selection ($w=1$) & 71.5 \\
\textbf{1D Temporal NMS Spans} & \textbf{73.1} \\
\bottomrule
\end{tabular}
}
\end{table}

\subsubsection{Analysis of Decoupled Audio-Latent Optimization.}
We investigate the impact of different optimization targets within the DALO module, with results summarized in Table~\ref{tab:ablation_dalo}. 
Compared to the unoptimized baseline, updating either the audio patch embeddings $\mathcal{P}$ or the thought tokens $\mathcal{T}$ individually yields clear performance gains.  
 By jointly updating both $\mathcal{T}$ and $\mathcal{P}$, our method achieves the highest accuracy and a significantly elevated final reward. This confirms a strong synergistic effect: allowing acoustic representations and latent reasoning trajectories to co-evolve during test-time search enables the model to escape local optima and discover more reliable multimodal alignment.

\begin{table}[h]
\centering
\caption{Ablation on DALO optimization Configuration on MMAU.}
\label{tab:ablation_dalo}
\resizebox{\linewidth}{!}{
\begin{tabular}{llccc}
\toprule
\textbf{Configuration} & \textbf{Acc (\%)} & \textbf{Final Reward $\mathcal{R}^*$} $\uparrow$\\
\midrule
Unoptimized ($M=0$) & 70.2 & -- \\
Patch Embeddings Only ($\mathcal{P}$) & 71.4 & 4.731 \\
Thought Tokens Only ($\mathcal{T}$) & 70.5 & 4.740 \\
\textbf{Joint Update ($\mathcal{T} + \mathcal{P}$)} & \textbf{73.1} & \textbf{5.239} \\
\bottomrule
\end{tabular}
}
\end{table}

\subsection{In-Depth Analytical Studies}

\begin{figure}[h]
    \centering
    \includegraphics[width=\linewidth]{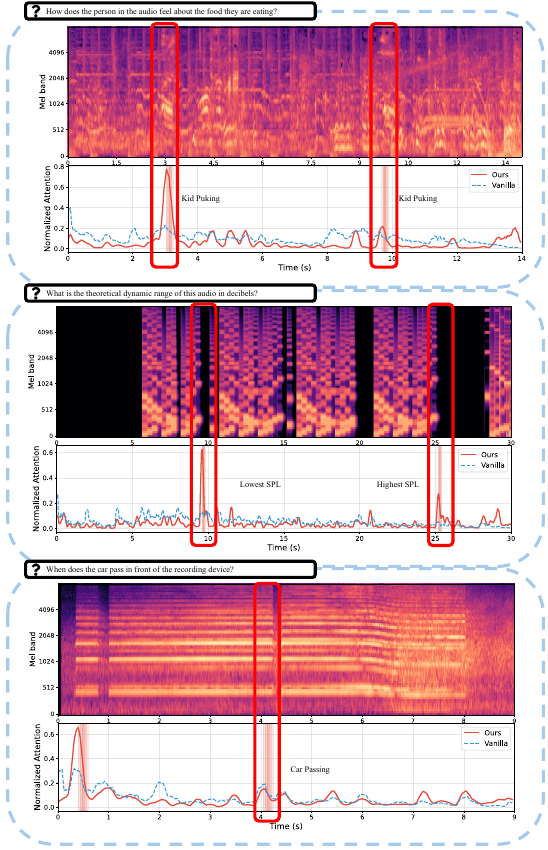} 
    \caption{Qualitative comparison of audio grounding behavior.}    
    \label{fig:latent-behavior}
\end{figure}

\paragraph{Audio Grounding Analysis.}
To qualitatively evaluate our model's grounding capability, we visualize the time-aligned attention distributions across diverse audio reasoning tasks, as shown in Figure~\ref{fig:latent-behavior}. Across all test cases, the Vanilla baseline exhibits a diffuse attention profile that frequently spikes around task-irrelevant background noise and silent intervals. In contrast, our approach dynamically focuses its attention on precise acoustic events critical for answering the query.

\paragraph{Latent Behavior Analysis.} To quantify the distributional shift induced by REINFORCE-based optimization, we measure the deviation of learnable patch embeddings from their corresponding ground-truth audio representations. For each sample, we record three quantities at every span index: (1) $d(P^{(0)}, P)$, the cosine distance between the initial raw-audio embedding and the final optimized patch, capturing the pure RL drift; (2) $d(P^{(0)}, Z_{C^*})$, the distance from the initial embedding to the reinstated raw-audio representation at the optimal NMS positions $C^*$, reflecting the initial offset from the true manifold; and (3) $d(P^, Z_{C^*})$, the distance from the drifted patch to the ground-truth audio, measuring the post-optimization deviation.

\begin{figure}[h]
    \centering
    \includegraphics[width=\linewidth]{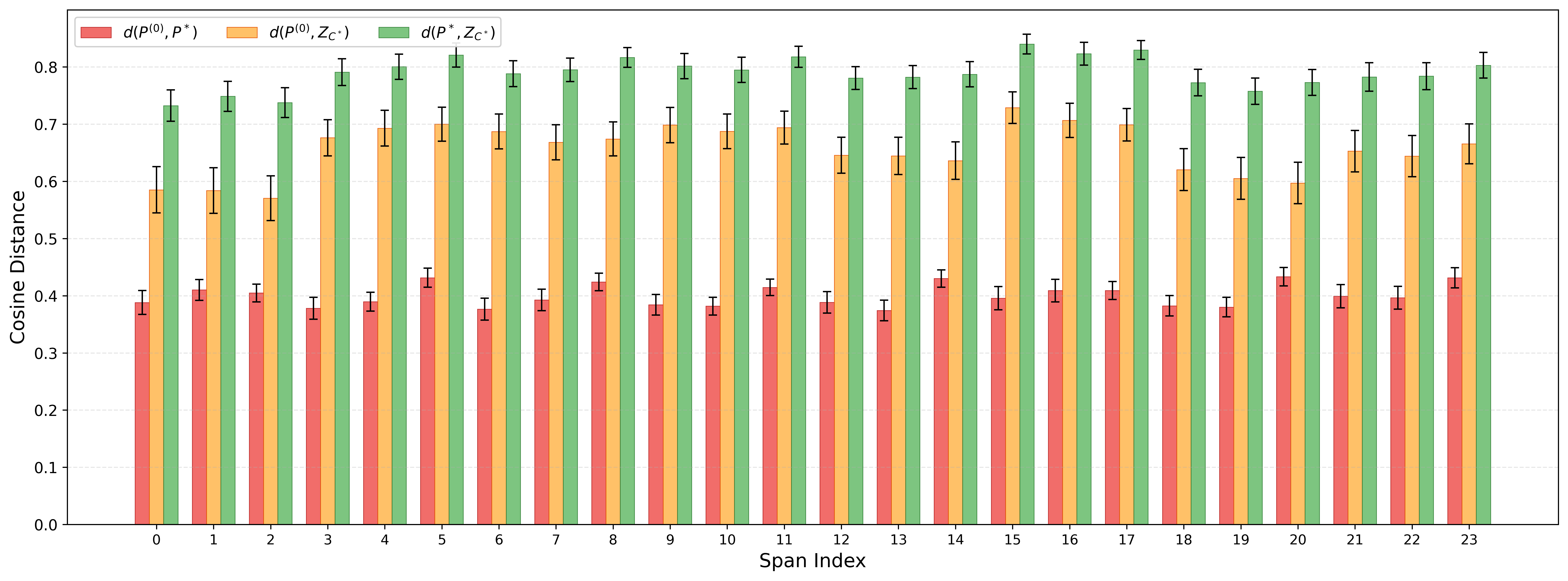} 
    \caption{Latent Behavior Analysis across span indices. Comparison of cosine distances $d(P^{(0)}, P^*)$, $d(P^{(0)}, Z_{C^*})$, and $d(P^*, Z_{C^*})$.}
    \label{fig:latent-drift}
\end{figure}

Figure~\ref{fig:latent-drift} reports these three distances averaged across samples over all span indices. Two observations stand out. First, $d(P^*, Z_{C^*})$ consistently exceeds $d(P^{(0)}, Z_{C^*})$ at every span position. This confirms that REINFORCE optimization systematically pushes learned patch embeddings away from the valid acoustic manifold — even when the patches are initialized from in-distribution audio features. Second, $d(P^{(0)}, P^*)$ demonstrates that the optimization process induces substantial representational change, moving representations further from the manifold of real audio. Audio Embedding Reinstatement resolves this drift by replacing $P^*$ with $Z_{C^*}$ to reduce the deviation. It restores the patches to their authentic acoustic distribution without sacrificing the optimal positions.

\paragraph{Efficiency Trade-off.}
Table~\ref{tab:efficiency_tradeoff} illustrates the trade-off between reasoning accuracy and inference latency across representative methods on the MMAU benchmark. While explicit textual reasoning (CCoT) enhances accuracy , it incurs a severe latency penalty due to verbose autoregressive decoding. Conversely, explicit interleaved reasoning (ICoT) and dynamic latent patch injection (DMLR) suffer from severe accuracy degradation. Remarkably, our \textsc{DeALR} achieves a state-of-the-art accuracy of $73.1\%$ requiring only $3.45~s$ per query while simultaneously improving accuracy. This confirms that \textsc{DeALR} achieves an optimal balance between reasoning effectiveness and computational efficiency.

\begin{table}[t]
\centering
\caption{Comparison of question-answering accuracy, inference latency, and computational efficiency across different reasoning methods on the MMAU benchmark. Best and second-best results are in bold and underlined, respectively.}
\label{tab:efficiency_tradeoff}
\setlength{\tabcolsep}{5.5pt}
\resizebox{\linewidth}{!}{
\begin{tabular}{lcccc}
\toprule
\textbf{Method} & \textbf{Acc (\%) $\uparrow$} & \textbf{$\Delta$ Acc} & \textbf{Latency (s) $\downarrow$}  \\
\midrule
Vanilla Baseline & 70.2 & -- & \textbf{1.05}  \\
CCoT & \underline{72.7} & \underline{+2.5} & 10.10 \\
ICoT & 38.0 & -32.2 & 4.43 \\
LTPO & 71.5 & +1.3 & \underline{3.07}  \\
DMLR & 67.3 & -2.9 & 3.77\\
\midrule
\textbf{\textsc{DeALR} (Ours)} & \textbf{73.1} & \textbf{2.9} & 3.45  \\
\bottomrule
\end{tabular}
}
\end{table}

\section{Conclusion}

In this paper, we present \textsc{DeALR}, a novel test-time latent reasoning framework designed to overcome the text-mediated reasoning bottleneck and modality attention imbalance in Large Audio-Language Models. To actively ground the model's deliberation on acoustic evidence, \textsc{DeALR} introduces Span-Level Acoustic Backtracking, which leverages initial thought attention to isolate contiguous, highly salient acoustic spans. Furthermore, we propose Decoupled Audio-Latent Optimization (\textsc{DALO}), a test-time policy search mechanism that jointly optimizes latent thought vectors and acoustic patch embeddings under predictive confidence rewards. Crucially, \textsc{DALO} incorporates an Audio Embedding Reinstatement step to re-inject raw acoustic embeddings during final decoding, effectively reducing optimization-induced distribution drift. Comprehensive evaluations on MMAR, MMAU, and AudioMCQ benchmarks demonstrate that \textsc{DeALR} consistently achieves state-of-the-art performance among the evaluated test-time reasoning methods, restoring modality attention balance, improving confidence calibration, and establishing an effective paradigm for test-time multimodal latent exploration.







\bibliography{aaai2027}



\end{document}